\documentclass[12pt,a4paper]{article}%
\usepackage{amsfonts}
\usepackage{graphicx}
\usepackage{amsmath}
\usepackage{amssymb}
\usepackage{fancyhdr}%
\begin{document}

\author{Piroska D\"{o}m\"{o}t\"{o}r\thanks{Domotor.Piroska@stud.u-szeged.hu}
\ and
Mih\'{a}ly G. Benedict\thanks{benedict@physx.u-szeged.hu}\\
\emph{Department of Theoretical Physics, University of Szeged, Hungary}\\}
\title{Global entanglement and coherent states in an $N$-partite system }
\date{}
\maketitle

\begin{abstract}
We consider a quantum system consisting of N parts, each of which is a
\textquotedblleft quKit\textquotedblright\ described by a $K$ dimensional
Hilbert space. We prove that in the symmetric subspace, $\mathbb{S}$, a pure
state is not globally entangled, if and only if it is a coherent state. It is
also shown that in the orthogonal complement $\mathbb{S}_{\bot}$ all states
are globally entangled.

\end{abstract}

\section{Introduction}

The property of entanglement of composite systems is considered to be the most
pronounced difference between classical and quantum systems, and it has become
especially in the center of interest since the realization that it can serve
as a resource for quantum computation and communication \cite{NC00}. The
definition of entanglement is easily understood for pure states, while the
full characterization of the entanglement properties of mixed states is a
difficult and still unsolved mathematical problem \cite{HHHH07}. We note,
however, that from the point of view of physical applications, \emph{pure}
state entanglement does not seem to be less important, as quantum algorithms
and communication protocols are in principle based on pure states \cite{NC00}.

Another important concept widely used in quantum mechanics is the notion of
\emph{coherent}, or quasiclassical states \ Introduced first for the harmonic
oscillator \cite{G63b} the construction has been extended to several other
systems, and the general treatment of coherent states can be found in the
books \cite{KS85,P86,AAG00}. Oscillator coherent states are exceptional in
many respects, they yield the classical results for the expectation values of
all the pertinent operators, and they are pointer states for a particle moving
in one dimension in a disspative environment \cite{ZHP93}. A similar result
holds for a collection of spins \cite{FCB01a}. From the point of view of the
present work we quote the paper \cite{ACGT72}, where coherent states of
several two level atoms (or qubits) are considered, and the review
\cite{ZFG90}, where coherent states are considered in a more general context
of Lie groups and algebras. We shall use the methods summarized in that work,
but our presentation should be understood in terms of the usual properties of
Hilbert space vectors.

In a previous paper \cite{DB08} we considered the question of entanglement of
$N$ qubits, while here we present the extension of the problem to the case of
\textquotedblleft\emph{quKits}\textquotedblright, where the constituent
systems are of arbitrary finite dimension, $K$. We shall prove the statement
that a multipartite pure quKit state in the symmetric subspace is not
entangled if and only if it is a coherent state, while in the subspace
orthogonal to the symmetric one, all states are entangled. A related property
between coherent states and entanglement was pointed out earlier in
\cite{BMR98} for a two-partite system. We also cite a recent work on symmetric
states and entanglement \cite{ISTY08}.

\section{Multipartite system of quKits, and operators on it}

We consider an $N$ partite system where each constituent is described by a $K$
dimensional Hilbert space, $\mathfrak{H}\equiv%
\mathbb{C}
^{K}$ \ ($j=1,\ldots N$) \ the elements of which can be called quKits. A
general pure state in the tensor product $\mathfrak{H}^{\otimes N}%
\equiv\mathbb{C}^{K^{N}}$can be given in a product basis $%
{\textstyle\bigotimes\limits_{j=1}^{N}}
\left\vert k_{j}\right\rangle \equiv\left\vert k_{1}\right\rangle \left\vert
k_{2}\right\rangle \ldots\left\vert k_{N}\right\rangle $, formed by some fixed
orthonormal bases from each subsystem $\{\left\vert k\right\rangle
\}=(\left\vert 1\right\rangle ,\left\vert 2\right\rangle ,\ldots\left\vert
K\right\rangle ).$ Then a general pure state in $\mathbb{C}^{K^{N}}$ is%
\begin{equation}
\left\vert \psi\right\rangle =\sum_{k_{1},k_{1},k_{2}\ldots k_{N}=1}%
^{K}c(k_{1}k_{2}\ldots k_{N})\left\vert k_{1}\right\rangle \left\vert
k_{2}\right\rangle \ldots\left\vert k_{N}\right\rangle
\end{equation}
\newline where $c(k_{1}k_{2}\ldots k_{N})$ are complex constants indexed by
the $K^{N}$ possible values of the set in the parentheses, and normalized to
1: $\sum_{k_{1},k_{1},k_{2}\ldots k_{N}=1}^{K}\left\vert c(k_{1}k_{2}\ldots
k_{N})\right\vert ^{2}=1.$ Operators in the tensor product space are built
from those in the individual $\mathfrak{H}$-s, and we shall use a standard set
of operators (an operator basis) in each subspace. The raising and lowering
operators are defined as $E_{ji}^{\dagger}=E_{ij}=\left\vert i\right\rangle
\!\left\langle j\right\vert $ and $E_{ji}=\left\vert j\right\rangle
\!\left\langle i\right\vert $ for all $i>j$, their number is $K(K-1)/2.$ They
are traceless operators acting on the chosen basis elements in the following
way: $E_{ij}\left\vert k\right\rangle =\delta_{kj}\left\vert i\right\rangle $
and $E_{ji}\left\vert k\right\rangle =\delta_{ki}\left\vert j\right\rangle $.
Their commutator is easily found to be $[E_{ij},E_{ij}^{\dagger}]=\left\vert
i\right\rangle \!\left\langle i\right\vert -\left\vert j\right\rangle
\!\left\langle j\right\vert =:H_{ij}$ which is also traceless and diagonal in
the selected basis. In the $K$ dimensional space there are only $K-1$ linearly
independent operators (apart from the identity) that can be diagonalized
simultaneously, and we can choose them to be the $H_{i+1,i}.$ The operators
$E_{ij},E_{ij}^{\dagger},H_{i+1,i}$ form a basis, called the Cartan-Weyl basis
in the (complex) space of the $K$ dimensional traceless operators, spanning
the Lie algebra sl$(K,\mathbb{C})$ \cite{G74}. In the $K=2$ case (qubits) they
are equivalent to the Pauli operators: $H_{21}\simeq\sigma_{3},$
$E_{12}^{\dagger}=E_{21}\simeq\sigma^{+}=(\sigma_{1}+i\sigma_{2})/2,$
$E_{12}\simeq\sigma^{-}=(\sigma_{1}-i\sigma_{2})/2$.

The corresponding collective operators in the $%
{\textstyle\bigotimes\limits_{j=1}^{N}}
\mathfrak{H}^{j}$ product space are defined as
\begin{equation}
\tilde{X}=\sum_{n=1}^{N}X^{\left(  n\right)  },\quad\text{where\qquad
\ }X^{\left(  n\right)  }=\underset{1}{\mathbf{1}}\otimes\underset
{2}{\mathbf{1}}\otimes\cdots\otimes\underset{n}{X}\otimes\cdots\otimes
\underset{N}{\mathbf{1}} \label{symop}%
\end{equation}
acts nontrivially only on the $n$th subsystem.

\section{A condition of unentanglement\label{unen}}

A pure $N$ partite state is not entangled by definition, if it is a product
state. For a system consisting of $N$ identical \emph{quKits} we shall use a
formal inductive method in order to decide if a pure state is entangled or
not. This is a generalization of a condition given by Meyer and Wallach
\cite{MW02}

Any pure state $\left\vert \psi\right\rangle $ $\in$ $\mathbb{C}^{K^{N}}$,
which is expanded in the standard basis $\{\left\vert k\right\rangle
\}^{\otimes N},$ can be decomposed for each $n=1,2\ldots N$ quKit as
\begin{equation}
\left\vert \psi\right\rangle =\left\vert 1\right\rangle _{n}\otimes\left\vert
u_{n}^{1}\right\rangle +\left\vert 2\right\rangle _{n}\otimes{\left\vert
u_{n}^{2}\right\rangle +}\left\vert 3\right\rangle _{n}\otimes{\left\vert
u_{n}^{3}\right\rangle +\cdots+}\left\vert K\right\rangle _{n}\otimes
{\left\vert u_{n}^{K}\right\rangle }=\sum_{k=1}^{K}\left\vert k\right\rangle
_{n}\otimes\left\vert u_{n}^{k}\right\rangle , \label{deco}%
\end{equation}
where the kets $\left\vert k\right\rangle _{n}$ are the basis vectors in
$\mathfrak{H}^{n}$ and $\left\vert u_{n}^{k}\right\rangle $ $\left(
k=1,\cdots K\right)  $ are vectors in $\mathbb{C}^{K^{N-1}}$ which are not
normalized in general. Using the above decomposition (\ref{deco}), it can be
simply proven that $\left\vert \psi\right\rangle $ is a product state
($\left\vert \psi\right\rangle $ is not entangled), if and only if the vectors
$\left\vert u_{n}^{k}\right\rangle $ are parallel to each other ($\left\vert
u_{n}^{1}\right\rangle \ \Vert\ \left\vert u_{n}^{2}\right\rangle
\ \Vert\ \cdots\ \Vert\ \left\vert u_{n}^{K}\right\rangle $) \ for all
possible $n$.

In order to see this, first assume that $\left\vert \psi\right\rangle $ is a
product state, and therefore it can be written as
\begin{equation}
\left\vert \psi\right\rangle ={\bigotimes\limits_{n=1}^{N}}(a_{n}%
^{1}\left\vert 1\right\rangle _{n}+a_{n}^{2}\left\vert 2\right\rangle
_{n}+\ldots a_{n}^{K}\left\vert K\right\rangle _{n})={\bigotimes
\limits_{n=1}^{N}}\sum_{k=1}^{K}a_{n}^{k}\left\vert k\right\rangle _{n}
\label{prod}%
\end{equation}
with some $a_{n}^{k}$, obeying $\sum_{k=1}^{K}\left\vert a_{n}^{k}\right\vert
^{2}=1$ for each $n$. We can write this product state as%
\begin{align}
\left\vert \psi\right\rangle  &  =a_{n}^{1}\left\vert 1\right\rangle
_{n}\bigotimes\limits_{j\neq n}^{N}\sum_{k=1}^{K}a_{j}^{k}\left\vert
k\right\rangle _{j}+a_{n}^{2}\left\vert 2\right\rangle _{n}\bigotimes
\limits_{j\neq n}^{N}\sum_{k=1}^{K}a_{j}^{k}\left\vert k\right\rangle
_{j}\ldots+a_{n}^{K}\left\vert K\right\rangle _{n}\bigotimes\limits_{j\neq
n}^{N}\sum_{k=1}^{K}a_{j}^{k}\left\vert k\right\rangle _{j}=\nonumber\\
&  =\sum_{k=1}^{K}\left\vert k\right\rangle _{n}\left\{  a_{n}^{k}%
\bigotimes\limits_{j\neq n}^{N}\sum_{k=1}^{K}a_{j}^{k}\left\vert
k\right\rangle _{j}\right\}
\end{align}
for each $n.$ We have then $\left\vert u_{n}^{k}\right\rangle =a_{n}%
^{k}\bigotimes\limits_{j\neq n}^{N}\sum_{k=1}^{K}a_{j}^{k}\left\vert
k\right\rangle _{j}$, which means that all the $\left\vert u_{n}%
^{k}\right\rangle $-s are parallel for a fixed $n$ in the decomposition
(\ref{deco}).

Consider now the reverse statement and assume parallellity of $\left\vert
u_{n}^{k}\right\rangle $ -s for each fixed $n=1,2\ldots N$ in (\ref{deco}). In
other words we assume that $\left\vert u_{n}^{k}\right\rangle =\beta_{n}%
^{k}\left\vert u_{n}^{1}\right\rangle $ for $k=2,\cdots K$ with some
coefficients $\beta_{n}^{k}\in%
\mathbb{C}
$.

Then $\left\vert \psi\right\rangle $ can be written in the form:%
\begin{equation}
\left\vert \psi\right\rangle =(1+\sum_{k=2}^{K}\left\vert \beta_{n}%
^{k}\right\vert ^{2})^{-1/2}\left(  \left\vert 1\right\rangle _{n}+\sum
_{k=2}^{K}\beta_{n}^{k}\left\vert k\right\rangle _{n}\right)  \otimes
\left\vert \tilde{u}_{n}^{1}\right\rangle \ \quad\forall\ n, \label{parallel}%
\end{equation}
and we prove by induction that $\left\vert \psi\right\rangle $ is a product
state (Here the $N-1$ \emph{quKit} states $\left\vert \tilde{u}_{n}%
^{k}\right\rangle =\left\vert u_{n}^{k}\right\rangle /\sqrt{\langle u_{n}%
^{k}|u_{n}^{k}\rangle}$ are normalized.) For $N=2$ this is obviously true,
because then\ $\left\vert \psi\right\rangle =(1+\sum_{k=2}^{K}\left\vert
\beta^{k}\right\vert ^{2})^{-1/2}\left(  \left\vert 1\right\rangle _{1}%
+\sum_{k=2}^{K}\beta_{1}^{k}\left\vert k\right\rangle _{1}\right)
\otimes\left\vert \tilde{u}_{1}^{1}\right\rangle $ and $\left\vert \tilde
{u}_{1}^{1}\right\rangle $ is a one \emph{quKit} state.

Suppose now that the statement is true for\ a system of $N-1$ \emph{quKits },
and proceed to $N$. We first use the decomposition (\ref{parallel}) with
respect to the $i$-th \emph{quKit}, where $\left\vert \tilde{u}^{i}%
\right\rangle \in\mathbb{C}^{K^{N-1}}$ is now an $N-1$ \emph{quKit} state.
Decompose $\left\vert \tilde{u}^{i}\right\rangle $ further, with respect to
the $j$-th \emph{quKit}:
\begin{align}
\left\vert \psi\right\rangle  &  =\frac{\left(  \left\vert 1\right\rangle
_{i}+\sum_{k=2}^{K}\beta_{i}^{k}\left\vert k\right\rangle _{i}\right)  }%
{\sqrt{1+\sum_{k=2}^{K}\left\vert \beta_{i}^{k}\right\vert ^{2}}}%
\otimes\left\vert \tilde{u}_{i}^{1}\right\rangle =\frac{\left(  \left\vert
1\right\rangle _{i}+\sum_{k=2}^{K}\beta_{i}^{k}\left\vert k\right\rangle
_{i}\right)  }{\sqrt{1+\sum_{k=2}^{K}\left\vert \beta_{i}^{k}\right\vert ^{2}%
}}\otimes\left(  \sum_{k=1}^{K}\left\vert k\right\rangle _{j}\otimes\left\vert
u_{ij}^{k}\right\rangle \right) \nonumber\\
&  =\sum_{k=1}^{K}\left\vert k\right\rangle _{j}\otimes\left(  \frac{\left(
\left\vert 1\right\rangle _{i}+\sum_{k=2}^{K}\beta_{i}^{k}\left\vert
k\right\rangle _{i}\right)  }{\sqrt{1+\sum_{k=2}^{K}\left\vert \beta_{i}%
^{k}\right\vert ^{2}}}\otimes\left\vert u_{ij}^{k}\right\rangle \right)
\end{align}
and compare this with
\begin{equation}
\left\vert \psi\right\rangle =\frac{\left(  \left\vert 1\right\rangle
_{j}+\sum_{k=2}^{K}\beta_{j}^{k}\left\vert k\right\rangle _{j}\right)  }%
{\sqrt{1+\sum_{k=2}^{K}\left\vert \beta_{j}^{k}\right\vert ^{2}}}%
\otimes\left\vert \tilde{u}_{j}^{1}\right\rangle .
\end{equation}
As a result we get
\begin{align}
\frac{\left\vert \tilde{u}_{j}^{1}\right\rangle }{\sqrt{1+\sum_{k=2}%
^{K}\left\vert \beta_{j}^{k}\right\vert ^{2}}}  &  {=}\frac{\left(  \left\vert
1\right\rangle _{i}+\sum_{k=2}^{K}\beta_{i}^{k}\left\vert k\right\rangle
_{i}\right)  }{\sqrt{1+\sum_{k=2}^{K}\left\vert \beta_{i}^{k}\right\vert ^{2}%
}}\otimes\left\vert u_{ij}^{1}\right\rangle \\
\beta_{j}^{k}\frac{\left\vert \tilde{u}_{j}^{1}\right\rangle }{\sqrt
{1+\sum_{k=2}^{K}\left\vert \beta_{j}^{k}\right\vert ^{2}}}  &  =\frac{\left(
\left\vert 1\right\rangle _{i}+\sum_{k=2}^{K}\beta_{i}^{k}\left\vert
k\right\rangle _{i}\right)  }{\sqrt{1+\sum_{k=2}^{K}\left\vert \beta_{i}%
^{k}\right\vert ^{2}}}\otimes\left\vert u_{ij}^{k}\right\rangle
\ \ \ \ \text{for all }k=2,\cdots K
\end{align}
which implies that $\beta_{j}^{k}\left\vert u_{ij}^{1}\right\rangle
=\left\vert u_{ij}^{k}\right\rangle $. As by hypothesis $\left\vert \tilde
{u}_{i}^{1}\right\rangle $ is a product state, and according to $\left\vert
\psi\right\rangle =\left(  1+\sum_{k=2}^{K}\left\vert \beta_{i}^{k}\right\vert
^{2}\right)  ^{-\frac{1}{2}}\left(  \left\vert 1\right\rangle _{i}+\sum
_{k=2}^{K}\beta_{i}^{k}\left\vert k\right\rangle _{i}\right)  \otimes
\left\vert \tilde{u}_{i}^{1}\right\rangle $, the $N$ \emph{quKit} state
$\left\vert \psi\right\rangle $ is also a product state.

\section{The symmetric subspace $\mathbb{S}$ of $\mathbb{C}^{K^{N}}$ \ }

We use the standard procedure to construct the symmetric subspace. Consider a
state, where the number of subsystems in states $\left\vert 1\right\rangle ,$
$\left\vert 2\right\rangle ,$ $\left\vert 3\right\rangle ;\cdots$ $\left\vert
K\right\rangle ;$ are $n_{1},$ $n_{2},$ $n_{3},$ $\cdots n_{K},$ respectively,
with $n_{k}\in%
\mathbb{N}
$. We have of course $\sum_{k=1}^{K}n_{k}=N$. Starting from a specific
\emph{non}symmetric state where we do know \emph{which} of the subsystems are
in the specified states, e.g. from
\begin{equation}
\left\vert \varphi(n_{1},\ldots,n_{K})\right\rangle :=\underbrace{\left\vert
1\right\rangle \left\vert 1\right\rangle \ldots\left\vert 1\right\rangle
}_{n_{1}}\underbrace{\left\vert 2\right\rangle \ldots\left\vert 2\right\rangle
}_{n_{2}}\left\vert 3\right\rangle \ldots\ldots\left\vert K-1\right\rangle
\underbrace{\left\vert K\right\rangle \ldots\left\vert K\right\rangle }%
_{n_{K}} \label{fiqu}%
\end{equation}
we get a symmetric state when applying the symmetrizer $\mathcal{S}$:
\begin{equation}
\mathcal{S}\left\vert \varphi(n_{1},\ldots,n_{K})\right\rangle =\mathcal{C}%
\sum_{\nu}P_{\nu}\left\vert \varphi(n_{1},\ldots,n_{K})\right\rangle
=:\left\vert n_{1},n_{2}\ldots n_{K}\right\rangle _{S} \label{symstate}%
\end{equation}
where $P_{\nu}$ runs over all the permutations of the $N$ subsystems, and
$\mathcal{C}$ is an appropriate normalization constant. This state also shares
the property, that \emph{the number of }subsystems in the specified state
$\left\vert k\right\rangle $ is $n_{k}$, but we do \emph{not} know which of
the subsystems is in a given basis state, as all such possibilities have the
same amplitude.

The number of the possible different symmetric states is easily obtained to
be\
\begin{equation}
\binom{N+K-1}{N}=\dim\mathbb{S} \label{dim}%
\end{equation}
as it is well known.

Alternatively we can get these symmetric states by applying raising operators
to the specific state $\left\vert \varphi(n_{1}=N,n_{2}=0,\ldots
,n_{K}=0)\right\rangle =\left\vert \varphi(N\ ,0,\ldots,0)\right\rangle
=\left\vert 1\right\rangle _{1}\otimes\left\vert 1\right\rangle _{2}%
\otimes\cdots\otimes\left\vert 1\right\rangle _{N}\equiv\left\vert
1,1\cdots1\right\rangle ,$ which is obviously symmetric. The state $\left\vert
1,1\cdots,1\right\rangle $ is the only one having the annihilation property
$\tilde{E}_{ji}\left\vert 1,1\cdots,1\right\rangle =0$ for all $i>j$, therfore
it is the so called lowest weight state in the representation theory of Lie
algebras. We apply consecutively the different powers of the collective
raising operators, $\tilde{E}_{ji}^{\dagger}=$ $\tilde{E}_{ij}=\sum_{n=1}%
^{N}E_{ij}^{\left(  n\right)  },$ (see definition (\ref{symop})), and get%
\begin{equation}
\frac{1}{_{n_{1}!n_{2}!\cdots n_{K}!}}\left(  \frac{N!}{_{n_{1}!n_{2}!\cdots
n_{K}!}}\right)  ^{-1/2}\left(  \mathbf{1}\right)  ^{n_{1}}\cdot\tilde
{E}_{2,1}^{n_{2}}\cdot\tilde{E}_{3,1}^{n_{3}}\cdot\cdots\cdot\tilde{E}%
_{K,1}^{n_{K}}\left\vert 1,1,\cdots1\right\rangle =\left\vert n_{1}%
,n_{2}\ldots n_{K}\right\rangle _{S}%
\end{equation}
with $\sum_{k=1}^{K}n_{k}=N.$ The initial state, as well as all the applied
operators according to (\ref{symop}) are symmetric, so the resulting state is
also symmetric. Two states of the type $\left\vert \varphi(n_{1},\ldots
n_{K})\right\rangle $ where the series of numbers $n_{1},\ldots n_{K}$ are not
identical are obviously orthogonal, and this property is inherited by their
symmetrized versions $\left\vert n_{1},n_{2}\ldots n_{K}\right\rangle _{S}$,
too. These states are eigenstates of the self-adjoint operators $\tilde
{H}_{i1}$:%
\begin{equation}
\tilde{H}_{i1}\left\vert n_{1},n_{2}\ldots n_{K}\right\rangle _{S}%
=(n_{i}-n_{1})\left\vert n_{1},n_{2}\ldots n_{K}\right\rangle _{S}%
\end{equation}
therefore a different set of $n_{i}$-s means orthogonal states.

In what follows, it will be expedient to use unnormalized versions of the
states $\left\vert n_{1},n_{2}\ldots n_{K}\right\rangle _{S}$:%
\begin{equation}%
\genfrac{\vert}{\rangle}{0pt}{}{N}{n_{1},n_{2},\cdots,n_{K}}%
:=\left(  \frac{N!}{_{n_{1}!n_{2}!\cdots n_{K}!}}\right)  ^{1/2}\left\vert
n_{1},n_{2}\ldots n_{K}\right\rangle _{S} \label{unnor}%
\end{equation}
For example: .$%
\genfrac{\vert}{\rangle}{0pt}{}{3}{1,2,0}%
=\sqrt{3}\left\vert 122\right\rangle _{S}=\left\vert 122\right\rangle
+\left\vert 212\right\rangle +\left\vert 221\right\rangle $ or $%
\genfrac{\vert}{\rangle}{0pt}{}{4}{3,0,0,1}%
=\sqrt{4}\left\vert 1114\right\rangle _{S}=\left\vert 1114\right\rangle
+\left\vert 1141\right\rangle +\left\vert 1411\right\rangle +\left\vert
4111\right\rangle .$

We note that we can get $%
\genfrac{\vert}{\rangle}{0pt}{}{N}{n_{1},n_{2},\cdots,n_{K}}%
$ from the $\left\vert \varphi(n_{1},\ldots,n_{K})\right\rangle $ states in a
direct way. To this end factorise the full permutation group $\mathcal{S}_{N}$
of the $N$ subsystems with the maximum stability group of the state
$\left\vert \varphi(n_{1},\ldots n_{K})\right\rangle $: $\mathcal{S}%
_{N}\diagup\mathcal{G},$ where \newline$\mathcal{G}=\left\{  P_{\nu}\in
S_{N}\mid P_{\nu}\left\vert \varphi(n_{1},\ldots n_{K})\right\rangle
=\left\vert \varphi(n_{1},\ldots n_{K})\right\rangle \right\}  $ is the group
of all permutations that leave $\left\vert \varphi\right\rangle $ invariant.
(The permutations in $\mathcal{G}$ only rearrange those subsystems, which are
in the identical\ states $\left\vert k\right\rangle $ as given by $\left\vert
\varphi(n_{1},\ldots n_{K})\right\rangle .)$ Applying the sum of all different
representing elements of the coset space we get the unnormalized symmetric
states:%
\begin{equation}
\sum_{\tilde{P}_{\nu}\in\mathcal{S}_{N}\diagup\mathcal{G}}\tilde{P}_{\nu
}\left\vert \varphi(n_{1},\ldots,n_{K})\right\rangle =:%
\genfrac{\vert}{\rangle}{0pt}{}{N}{n_{1},n_{2},\cdots,n_{K}}%
\end{equation}

\section{Generalized coherent states}

We recall the definition of coherent states for a general quantum system
\cite{P86,ZFG90}, applied here for our $N$ quKit states. The construction
follows that of the oscillator coherent states, where a continuously
parametrized set of unitary displacement operators shifts\ the ground state to
a coherent state. Here we start again with the lowest weight state $\left\vert
1,1,\cdots1\right\rangle $, and apply the unitary displacement operators to
it:%
\begin{align}
\left\vert \vec{\eta}\right\rangle  &  :=U(\vec{\eta})\left\vert
1,1\cdots1\right\rangle =\exp\left(  \sum_{i>j}\eta_{ij}\tilde{E}_{ij}%
-\eta_{ij}^{\ast}\tilde{E}_{ij}^{\dag}\right)  \left\vert 1,1\cdots
1\right\rangle =\nonumber\\
&  =\exp\left(  \sum_{n=2}^{K}\eta_{n}\tilde{E}_{n,1}-\eta_{n}^{\ast}\tilde
{E}_{1,n}\right)  \left\vert 1,1,\cdots,1\right\rangle \label{disp}%
\end{align}
where $\vec{\eta}=\left\{  \eta_{ij}\in%
\mathbb{C}
,i>j\right\}  $ is an arbitrary set of $K(K-1)/2$ complex constants. The
summation in the expression of $U$ goes only for those operators, which do not
annihilate the lowest weight state. The exponential is an antihermitian
operator in $\mathbb{S}$, therefore the displacements $U(\vec{\eta})$ are
unitary transformations acting in $\mathbb{S}$.

With help of the generalized Baker-Campbell-Haussdorff formula \cite{ZFG90} we
get:%
\begin{align}
\left\vert \vec{\eta}\right\rangle  &  =\exp\left(  \sum_{i=2}^{K}\tau
_{i}\tilde{E}_{i,1}\right)  \exp\left(  \sum_{i=2}^{K}\gamma_{i}\tilde
{H}_{i,1}\right)  \exp\left(  -\sum_{i=2}^{K}\tau_{i}\tilde{E}_{1,i}\right)
\left\vert 1,1\cdots1\right\rangle =\nonumber\\
&  =\mathcal{N}\exp\left(  \sum_{i=2}^{K}\tau_{i}\tilde{E}_{i,1}\right)
\left\vert 1,1\cdots1\right\rangle \label{BCH}%
\end{align}
where $\mathcal{N}$\ is an appropriate normalization factor and $\tau_{i}\in%
\mathbb{C}
$ are some complex constants.

This follows from $\left[  \tilde{E}_{i,1},\tilde{E}_{1,j}\right]  =\tilde
{E}_{i,j}$ for $i\neq j,$ and from the facts that $\left\vert 1,1\cdots
1\right\rangle $ is the eigenstate of $\left[  \tilde{E}_{i,1},\tilde{E}%
_{1,i}\right]  =\tilde{H}_{i,1}$ and that all $\ \tilde{E}_{1,i}$-s annihilate
$\left\vert 1,1\cdots1\right\rangle $.

\section{The coherent states are product states, and these are the only ones}

Using (\ref{BCH}) we can show, that the states $\left\vert \vec{\eta
}\right\rangle $ can be factorized into products of quKits, as follows%
\begin{align}
\left\vert \vec{\eta}\right\rangle  &  =\mathcal{N}\exp\left(  \sum_{i=2}%
^{K}\tau_{i}\tilde{E}_{i,1}\right)  \left\vert 1,1\cdots1\right\rangle
=\mathcal{N}\exp\left(  \sum_{i=2}^{K}\tau_{i}\sum_{n=1}^{N}E_{i,1}^{\left(
n\right)  }\right)  \left\vert 1\right\rangle _{1}\otimes\left\vert
1\right\rangle _{2}\otimes\cdots\otimes\left\vert 1\right\rangle
_{N}=\nonumber\\
&  =\mathcal{N}\exp\left(  \sum_{i=2}^{K}\tau_{i}E_{i,1}^{\left(  1\right)
}\right)  \left\vert 1\right\rangle _{1}\otimes\exp\left(  \sum_{i=2}^{K}%
\tau_{i}E_{i,1}^{\left(  2\right)  }\right)  \left\vert 1\right\rangle
_{2}\otimes\cdots\otimes\exp\left(  \sum_{i=2}^{K}\tau_{i}E_{i,1}^{\left(
N\right)  }\right)  \left\vert 1\right\rangle _{N}=\nonumber\\
&  =\mathcal{N}%
{\displaystyle\bigotimes\limits_{n=1}^{N}}
\exp\left(  \sum_{i=2}^{K}\tau_{i}E_{i,1}^{\left(  n\right)  }\right)
\left\vert 1\right\rangle _{n}=\mathcal{N}%
{\displaystyle\bigotimes\limits_{n=1}^{N}}
\left(  \mathbf{1}+\sum_{i=2}^{K}\tau_{i}E_{i,1}^{\left(  n\right)  }\right)
\left\vert 1\right\rangle _{n}=\nonumber\\
&  =\mathcal{N}%
{\displaystyle\bigotimes\limits_{n=1}^{N}}
\left(  \left\vert 1\right\rangle +\tau_{2}\left\vert 2\right\rangle +\tau
_{3}\left\vert 3\right\rangle +\cdots+\tau_{K}\left\vert K\right\rangle
\right)  _{n} \label{cohprod}%
\end{align}
Here we have used that
\begin{equation}
{\exp\left(  \sum_{i=2}^{K}\tau_{i}E_{i,1}^{\left(  n\right)  }\right)
\left\vert 1\right\rangle _{n}={\sum\limits_{q=0}^{\infty}}\frac{1}{q!}\left(
\sum_{i=2}^{K}\tau_{i}E_{i,1}^{\left(  n\right)  }\right)  ^{q}\left\vert
1\right\rangle _{n}=\left(  \mathbf{1}+\sum_{i=2}^{K}\tau_{i}E_{i,1}^{\left(
n\right)  }\right)  \left\vert 1\right\rangle _{n}}%
\end{equation}
and exploited that $E_{i,1}^{\left(  n\right)  }$ -s "moves" only the
$\left\vert 1\right\rangle $ basis state, so $\left(  E_{j,1}\right)  ^{q_{j}%
}\left(  E_{i,1}\right)  ^{q_{i}}\left\vert 1\right\rangle _{n}=0$ for all $i$
and $j$ if $q_{i}+q_{j}\geqq2$.

Now we prove that in the symmetric subspace the notentangled states are
exactly the coherent states. We can write any pure state in \ $\mathbb{S}$ as
a linear combination of the unnormalized states introduced in (\ref{unnor}):
\begin{align}
\left\vert \psi\right\rangle  &  =\sum_{n_{1}+\cdots+n_{K}=N}C_{n_{1}%
,n_{2},\cdots,n_{K}}%
\genfrac{\vert}{\rangle}{0pt}{}{N}{n_{1},n_{2},\cdots,n_{K}}%
=\\
&  =\sum_{i_{2}=0}^{N}\sum_{i_{3}=0}^{N}\cdots\sum_{i_{K}=0}^{N}%
C_{N-i_{2}-i_{3}-\cdots-i_{K},i_{2},\cdots,i_{K}}%
\genfrac{\vert}{\rangle}{0pt}{}{N}{N-i_{2}-i_{3}-\cdots-i_{K},i_{2}%
,\cdots,i_{K}}%
.\nonumber
\end{align}
$\langle\psi|\psi\rangle=1$ then requires:%
\begin{equation}
\sum_{n_{1}+\cdots+n_{K}=N}^{N}\left\vert C_{n_{1},n_{2},\cdots,n_{K}%
}\right\vert ^{2}\frac{N!}{_{n_{1}!n_{2}!\cdots n_{K}!}}=1
\end{equation}

The states $%
\genfrac{\vert}{\rangle}{0pt}{}{N}{n_{1},n_{2},\cdots,n_{K}}%
$-s have the following property
\begin{gather}%
\genfrac{\vert}{\rangle}{0pt}{}{N}{n_{1},n_{2},\cdots,n_{K}}%
=\left\vert 1\right\rangle _{n}\otimes%
\genfrac{\vert}{\rangle}{0pt}{}{N-1}{n_{1}-1,n_{2},\cdots,n_{K}}%
+\left\vert 2\right\rangle _{n}\otimes%
\genfrac{\vert}{\rangle}{0pt}{}{N-1}{n_{1},n_{2}-1,\cdots,n_{K}}%
+\cdots\nonumber\\
\cdots+\left\vert K\right\rangle _{n}\otimes%
\genfrac{\vert}{\rangle}{0pt}{}{N-1}{n_{1},n_{2},\cdots,n_{K}-1}%
=\sum_{k=1}^{K}\left\vert k\right\rangle _{n}\otimes%
\genfrac{\vert}{\rangle}{0pt}{}{N-1}{n_{1},n_{2},\cdots,n_{k}-1,\cdots,n_{K}}%
, \label{combi}%
\end{gather}
where $%
\genfrac{\vert}{\rangle}{0pt}{}{N}{n_{1},n_{2},\cdots,n_{K}}%
=0$ by definition, if $n_{k}>N$ or $n_{k}<0$ for $k\in\left\{  1,2,\cdots
,K\right\}  .$The above decompositions, which correspond to the elementary
combinatoric identity $\frac{N!}{_{n_{1}!n_{2}!\cdots n_{K}!}}=\sum_{k=1}%
^{K}\frac{\left(  N-1\right)  !}{_{n_{1}!\cdots\left(  n_{k}-1\right)  !\cdots
n_{K}!}}$ are valid for any $n=1,\ldots N,$ as a consequence of the symmetry
of the states $%
\genfrac{\vert}{\rangle}{0pt}{}{N}{n_{1},n_{2},\cdots,n_{K}}%
$ with respect of permutations.

Then using (\ref{combi}), we get:%
\begin{align}
\left\vert \psi\right\rangle  &  =\sum_{n_{1}+\cdots+n_{K}=N}C_{n_{1}%
,n_{2},\cdots,n_{K}}%
\genfrac{\vert}{\rangle}{0pt}{}{N}{n_{1},n_{2},\cdots,n_{K}}%
\nonumber\\
&  =\sum_{n_{1}+\cdots+n_{K}=N}C_{n_{1},n_{2},\cdots,n_{K}}\sum_{k=1}%
^{K}\left\vert k\right\rangle _{n}\otimes%
\genfrac{\vert}{\rangle}{0pt}{}{N-1}{n_{1},n_{2},\cdots,n_{k}-1,\cdots,n_{K}}%
=\nonumber\\
&  =\sum_{i=1}^{K}\left(  \left\vert k\right\rangle _{n}\otimes\sum
_{n_{1}+\cdots+n_{K}=N}C_{n_{1},n_{2},\cdots,n_{K}}%
\genfrac{\vert}{\rangle}{0pt}{}{N-1}{n_{1},n_{2},\cdots,n_{k}-1,\cdots,n_{K}}%
\right)  .
\end{align}
As $\left\vert \psi\right\rangle $ is an arbitrary not entangled state, so
according to the parallellity criterion of section \ref{unen}, the following
$N-1$ partite states should be parallel to each other:
\begin{gather}
\sum_{n_{1}+\cdots+n_{K}=N}C_{n_{1},n_{2},\cdots,n_{K}}%
\genfrac{\vert}{\rangle}{0pt}{}{N-1}{n_{1}-1,n_{2},\cdots,n_{K}}%
\nonumber\\
\sum_{n_{1}+\cdots+n_{K}=N}C_{n_{1},n_{2},\cdots,n_{K}}%
\genfrac{\vert}{\rangle}{0pt}{}{N-1}{n_{1},n_{2}-1,\cdots,n_{K}}%
=\sum_{n_{1}+\cdots+n_{K}=N}C_{n_{1}-1,n_{2}+1,\cdots,n_{K}}%
\genfrac{\vert}{\rangle}{0pt}{}{N-1}{n_{1}-1,n_{2},\cdots,n_{K}}%
\nonumber\\
\vdots\nonumber\\
\sum_{n_{1}+\cdots+n_{K}=N}C_{n_{1},n_{2},\cdots,n_{K}}%
\genfrac{\vert}{\rangle}{0pt}{}{N-1}{n_{1},n_{2},\cdots,n_{K}-1}%
=\sum_{n_{1}+\cdots+n_{K}=N}C_{n_{1}-1,n_{2},\cdots,n_{K}+1}%
\genfrac{\vert}{\rangle}{0pt}{}{N-1}{n_{1}-1,n_{2},\cdots,n_{K}}%
\nonumber\\
\end{gather}
\newline By changing the summation indices in the right hand sides as
$n_{1}\rightarrow n_{1}^{\prime}=n_{1}+1$; $n_{k}\rightarrow n_{k}^{\prime
}=n_{k}-1,$ we see \ that the parallellity of these states requires :%
\begin{align}
\tau_{2}\cdot C_{n_{1},n_{2},\cdots,n_{K}} &  =C_{n_{1}-1,n_{2}+1,\cdots
,n_{K}}\nonumber\\
&  \vdots\label{partau}\\
\tau_{K}\cdot C_{n_{1},n_{2},\cdots,n_{K}} &  =C_{n_{1}-1,n_{2},\cdots
,n_{K}+1}\nonumber
\end{align}
where $\tau_{i}$ $\left(  i\in\left\{  2,\cdots,K\right\}  \right)  $.are
arbitrary complex numbers.

This means that $C_{N-i_{2}-i_{3}-\cdots-i_{K},i_{2},\cdots,i_{K}}$ must be of
the form:
\begin{equation}
C_{N-i_{2}-i_{3}-\cdots-i_{K},i_{2},\cdots,i_{K}}=\tau_{2}^{i_{2}}\cdot
\cdots\cdot\tau_{K}^{i_{K}}\cdot C_{N,0,\cdots,0}%
\end{equation}
And thus -- comparing with (\ref{cohprod}) -- we immediately see that
$\left\vert \psi\right\rangle $ is a coherent state%
\begin{gather}
\left\vert \psi\right\rangle =\sum_{i_{2}=0}^{N}\sum_{i_{3}=0}^{N}\cdots
\sum_{i_{K}=0}^{N}C_{N-i_{2}-i_{3}-\cdots-i_{K},i_{2},\cdots,i_{K}}%
\genfrac{\vert}{\rangle}{0pt}{}{N}{N-i_{2}-i_{3}-\cdots-i_{K},i_{2}%
,\cdots,i_{K}}%
\nonumber\\
=\sum_{i_{2}=0}^{N}\sum_{i_{3}=0}^{N}\cdots\sum_{i_{K}=0}^{N}\tau_{2}^{i_{2}%
}\cdot\cdots\cdot\tau_{K}^{i_{K}}\cdot C_{N,0,\cdots,0}%
\genfrac{\vert}{\rangle}{0pt}{}{N}{N-i_{2}-i_{3}-\cdots-i_{K},i_{2}%
,\cdots,i_{K}}%
\nonumber\\
=C_{N,0,\cdots,0}\cdot\left(  \left\vert 1\right\rangle +\tau_{2}\left\vert
2\right\rangle +\cdots+\tau_{K}\left\vert K\right\rangle \right)  ^{\otimes N}%
\end{gather}

\section{The vectors orthogonal to the symmetric subspace are all entangled}

We shall now consider vectors in $\mathbb{S}_{\bot}$ the orthogonal complement
of the symmetric space. We prove that all vectors in $\mathbb{S}_{\bot}$ are
globally entangled. To this end, assume to the contrary, that there exists a
vector $\left\vert \varphi\right\rangle \in$ $\mathbb{S}_{\bot}$which can be
written as a product:%
\begin{equation}
\left\vert \varphi\right\rangle ={\bigotimes\limits_{n=1}^{N}}(a_{n}%
^{1}\left\vert 1\right\rangle _{n}+a_{n}^{2}\left\vert 2\right\rangle
_{n}+\ldots a_{n}^{K}\left\vert K\right\rangle _{n})={\bigotimes
\limits_{n=1}^{N}}\sum_{k=1}^{K}a_{n}^{k}\left\vert k\right\rangle _{n}%
\end{equation}
with $\sum_{k=1}^{K}\left\vert a_{n}^{k}\right\vert ^{2}=1$ for each
$n=1.\cdots N$.

As $\left\vert \varphi\right\rangle \in$ $\mathbb{S}_{\bot}$ it cannot have a
nonzero projection onto an arbitrary symmetric state.

Consider first the projection of $\left\vert \varphi\right\rangle $onto the
symmetric state: $%
\genfrac{\vert}{\rangle}{0pt}{}{N}{N,0,\cdots,0}%
=\left\vert 1\right\rangle _{1}\otimes\left\vert 1\right\rangle _{2}%
\otimes\cdots\otimes\left\vert 1\right\rangle _{N}$. This must be zero, which
demands $\prod\limits_{n=1}^{N}a_{n}^{1}=0,$ so at least one of the $a_{n}%
^{1}$-s must be zero. Without loss of generality we may assume that the
vanishing coefficient is $a_{1}^{1}=0$.

Then $\left\vert \varphi\right\rangle =\left(  a_{1}^{2}\left\vert
2\right\rangle +a_{1}^{3}\left\vert 3\right\rangle +\cdots+a_{1}^{K}\left\vert
K\right\rangle \right)  _{1}%
{\displaystyle\bigotimes\limits_{n=2}^{N}}
\left(  a_{n}^{1}\left\vert 1\right\rangle +a_{n}^{2}\left\vert 2\right\rangle
+a_{n}^{3}\left\vert 3\right\rangle +\cdots+a_{n}^{K}\left\vert K\right\rangle
\right)  _{n}$

Consider now the projections onto the symmetric states with $n_{1}=N-1$, which
are $\frac{1}{\sqrt{N}}%
\genfrac{\vert}{\rangle}{0pt}{}{N}{N-1,1,\cdots,0}%
;\ \frac{1}{\sqrt{N}}%
\genfrac{\vert}{\rangle}{0pt}{}{N}{N-1,0,1,0\cdots,0}%
;\cdots\frac{1}{\sqrt{N}}%
\genfrac{\vert}{\rangle}{0pt}{}{N}{N-1,0,\cdots,1}%
$. As these projections must be zero again, we have $a_{1}^{k}\prod
\limits_{n=2}^{N}a_{n}^{1}=0,$ for all possible $k=2,\cdots,K$. All the
$a_{1}^{k}$ -s can not be zero, because then $\left\vert \varphi\right\rangle
$ would be zero, so $\prod\limits_{n=2}^{N}a_{n}^{1}=0$ and again at least one
of the $a_{n}^{1}$-s $\left(  n\neq1\right)  $ must be zero. Without loss of
generality we may assume that the vanishing coefficient is $a_{2}^{1}=0$.

Next we consider the symmetric states with $n_{1}=N-2$. The projections onto
these states again must be zero, which leads us finally to $\prod
\limits_{n=3}^{N}a_{n}^{1}=0$. Again one must be zero, say $a_{3.}^{1}$.
Continuing in this way with $n_{1}=N-3,\cdots,1,0$ we finally arrive to
$a_{1}^{1}=a_{2}^{1}=\cdots=a_{N}^{1}=0$.

If we carry out the same reasoning starting with the state corresponding to
$n_{2}=N$, we get $a_{1}^{2}=a_{2}^{2}=\cdots=a_{N}^{2}=0$, and continuing in
this way, with $n_{3}=\cdots=n_{K}=N,$ we get at the end that all the
$a_{n}^{k}$-s are zero, so $\left\vert \varphi\right\rangle =0$.

We arrived to a contradiction: the nonentangled $\left\vert \varphi
\right\rangle $ cannot be orthogonal to $\mathbb{S}$, or stated otherwise: all
elements of $\mathbb{S}_{\bot}$ are entangled.

In conclusion, our result shows that besides of other characteristics showing
quasiclassicity of coherent states, there exists an additional remarkable one:
they are the only nonentangled pure states in a\ symmetric multipartite system
of quKits. Moreover, all states in $\mathbb{S}_{\bot},$ in the orthogonal
complement of the symmetric subspace, are entangled. We note, however, that
the analogy between oscillator coherent states and the ones considered here
cannot be extended to the question of entanglement, as an oscillator mode is a
single system, while atomic coherent states are multipartite by their
definition, which is a necessary condition of entanglement.

The work was supported by the Hungarian Scientific Research Fund (OTKA)\ under
contract No: T48888.We thank L. Feh\'{e}r for useful discussions.

\end{document}